\documentclass[aps, twocolumn, superscriptaddress, amsmath,amssymb, longbibliography]{revtex4-1}
\usepackage{graphicx}% Include figure files
\usepackage{dcolumn}% Align table columns on decimal point
\usepackage{bm}% bold math
\usepackage{tabularx}
\usepackage{hyperref}
\usepackage[utf8]{inputenc}
\usepackage[T1]{fontenc}
\usepackage{mathptmx}
\usepackage{color, soul}
\usepackage{siunitx}
\usepackage{soul,xcolor}
\usepackage{lineno}
\usepackage{comment}
\graphicspath{ {./} }
\graphicspath{ {Figures/} }

\newcommand{\ba}{\begin{eqnarray}}
\newcommand{\ea}{\end{eqnarray}}

\begin{document}
\setstcolor{red}
\preprint{AIP/123-QED}

\title{Measuring kinetic inductance and superfluid stiffness of two-dimensional superconductors using high-quality transmission-line resonators}
%Alt(JH): Measuring kinetic inductance and superfluid stiffness using low-loss transmission-line resonators\\
%Alt: van der Waals superconductors for microwave quantum devices\\
%Alt: Quality van der Waals superconductor for quantum devices\\
% Force line breaks with \\
\author{Mary Kreidel}%
\affiliation{School of Engineering and
Applied Sciences, Harvard University, Cambridge, Massachusetts 02138, USA}
\author{Xuanjing Chu}
\affiliation{Department of Applied Physics and Applied Mathematics, Columbia University, New York, NY 10027, USA}
\author{Jesse Balgley}
\affiliation{Department of Mechanical Engineering, Columbia University, New York, NY 10027, USA}
\author{Abhinandan Antony}
\affiliation{Department of Mechanical Engineering, Columbia University, New York, NY 10027, USA}
\author{Nishchhal Verma}
\affiliation{Department of Physics, Columbia University, New York, NY 10027}
\author{Julian Ingham}
\affiliation{Department of Physics, Columbia University, New York, NY 10027}
\author{Leonardo Ranzani}
\affiliation{RTX BBN Technologies, Quantum Engineering and Computing Group, Cambridge, MA 02138, USA}
\author{Raquel Queiroz}
\affiliation{Department of Physics, Columbia University, New York, NY 10027}
\affiliation{Center for Computational Quantum Physics, Flatiron Institute, New York, NY 10010, USA}
\author{Robert M. Westervelt}
\affiliation{School of Engineering and
Applied Sciences, Harvard University, Cambridge, Massachusetts 02138, USA}
\affiliation{Department of Physics, Harvard University, Cambridge, Massachusetts 02138, USA}
\author{James Hone}
\affiliation{Department of Mechanical Engineering, Columbia University, New York, NY 10027, USA}
\author{Kin Chung Fong}%
\altaffiliation{Present address: k.fong@northeastern.edu, Northeastern University}
\affiliation{RTX BBN Technologies, Quantum Engineering and Computing Group, Cambridge, MA 02138, USA}
\date{\today}

\begin{abstract}
The discovery of van der Waals superconductors in recent years has generated a lot of excitement for their potentially novel pairing mechanisms. However, their typical atomic-scale thickness and micrometer-scale lateral dimensions impose severe challenges to investigations of pairing symmetry by conventional methods. In this report we demonstrate a new technique that employs high-quality-factor superconducting resonators to measure the kinetic inductance---up to a part per million---and loss of a van der Waals superconductor. We analyze the equivalent circuit model to extract the kinetic inductance, superfluid stiffness, penetration depth, and ratio of imaginary and real parts of the complex conductivity. We validate the technique by measuring aluminum and finding excellent agreement in both the zero-temperature superconducting gap as well as the complex conductivity data when compared with BCS theory. We then demonstrate the utility of the technique by measuring the kinetic inductance of multi-layered niobium diselenide and discuss the limits to the accuracy of our technique when the transition temperature of the sample, NbSe$_2$ at 7.06~K, approaches our Nb probe resonator at 8.59~K. Our method will be useful for practitioners in the growing fields of superconducting physics, materials science, and quantum sensing, as a means of characterizing superconducting circuit components and studying pairing mechanisms of the novel superconducting states which arise in layered 2D materials and heterostructures.
%The extracted values of zero-temperature gap energy match BCS theory for both the Al and the layered superconductor NbSe$_2$. We 
\end{abstract}
\maketitle

\section{Introduction}
The confluence of flat bands, quantum geometry and correlations gives rise to novel quantum phenomena in two-dimensional van der Waals (vdW) materials. For example, unconventional superconducting states have been observed in transition metal dichalcogenides (TMDs), as well as twisted and untwisted multilayers of graphene and TMDs \cite{Cao.2018,Lu.2019,Yankowitz.Dean.2019,Park.Jarillo-Herrero.2021,Hao.Kim.2021,Zhang.Nadj-Perge.2022,Wang.Dean.2020,Park.Jarillo-Herrero.2022,Saito.Iwasa.2016,Manzeli.2017,Andrei.Young.2021,Zhou.Young.2022,Zhang.Nadj-Perge.2023,Zhou.Young.2021,Ghiotto.Pasupathy.2021}. Accumulating evidence points to the unusual nature of both the superconducting pairing \cite{Oh.Yazdani.2021,Kim.Nadj-Perge.2022, Banerjee.Kim.2024} and the normal parent state of these systems \cite{Wong.Yazdani.2020,Zondiner.others.2020,Morissette.Li.2023}, demanding further experimental scrutiny. Superfluid stiffness is a fundamental quantity which gives insight into the nature of the pairing symmetry, and whose magnitude determines the relevance of phase fluctuations of the superconducting order parameter---of particular importance in strongly correlated superconductors characterized by low superfluid densities 
\cite{Emery.1995}. The superfluid stiffness is directly related to the London penetration depth $\lambda_L$, which characterizes the decay of an external magnetic field inside a superconductor due to the Meissner effect, and is traditionally a well-established quantity to study superconductivity \cite{Hardy.1993}. However, the typical atomic-scale thickness and micron-scale lateral dimensions of these van der Waals superconductors makes them challenging to measure with conventional methods, such as dielectric resonators \cite{Bae.2019}, tunnel-diode-driven resonators \cite{Fletcher.Giannetta.2007}, muon spin rotation \cite{Sonier.Brill.1997}, and the mutual inductance method \cite{Hetel.2007}.

Precise measurement of complex conductivity can act as a powerful tool to study the fundamental properties of superconductors \cite{Tinkham}. The reactive component of the conductivity is the kinetic inductance, which arises from the kinetic energy of Cooper pairs. Kinetic inductance can directly measure superfluid stiffness, and therefore, has the capability to reveal quantum geometric contributions to stiffness which are invisible at the level of band dispersion \cite{Tovmasyan.Huber.2016, Julku2016,Verma.Randeria.2021,Torma.2022}, and for which there only exists indirect evidence at present \cite{Tian.2023} in twisted graphene. Sensitive measurements of the superfluid stiffness can also be used to detect the weak supercurrent dependence of the penetration depth (known as the nonlinear Meissner effect \cite{Yip.1992,Xu.1995,Stojkovic.1995,Sauls.2022}), which offers highly constraining insights into the superconducting pairing symmetry. 
On the other hand, the resistive component of the complex conductivity reflects the energy dissipation due to normal-state carriers under the framework of the two-fluid model.
Finally, kinetic inductance is an important ingredient in superconducting quantum circuitry: it can be used to engineer qubits \cite{Winkel.2020}, to realize high-sensitivity photon detectors \cite{Day.2003}, and to parametrically amplify signals with performance constrained solely by quantum noise \cite{Eom.2012,Ranzani.2018}. 

In this paper, we describe a new technique for accurately determining the kinetic inductance of these thin, small superconductor sample components. Inspired by the use of superconducting resonators in the readout of qubit states in circuit quantum electrodynamics (cQED), our method is a mesoscopic hybrid setup in which the kinetic inductance of our sample component acts as a small perturbation to the probe resonator. We then extract the relevant sample component parameters from the change of the resonator frequency and quality factor. This contrasts with previous kinetic inductance measurements in which a 50~$\Omega$ resistor shunts out the resonator in order to impedance match to the measurement circuitry, thus reducing the quality ($Q$) factor and sacrificing measurement sensitivity \cite{Annunziata.2010, Singh.2018}. Our method shares the same fundamental mechanism as kinetic inductance detectors, which measure the suppression of resonance frequency due to the breaking of Cooper pairs in a superconducting resonator \cite{Visser.2014, Day.2003}. Already, this method has been applied to study the superfluid stiffness in superconductor-ferromagnet bilayers \cite{Bottcher.2023}, FeSe$_{1-x}$Te$_x$ \cite{Cherpak.2023}, and magic-angle twisted bilayer graphene \cite{wang.oliver.2024}, as well as metrological studies of the loss tangent of materials and the cleanliness of fabrication procedures to advance the development of superconducting quantum technology \cite{Sage.2011,Quintana.2014,Gruenhaupt.2018,McRae.2021,Antony.2021}.

Here, we will first introduce the theory of kinetic inductance and superfluid stiffness, before focusing on the relevant hybrid resonator circuit model, deriving its corresponding formulae, and validating our method with a well-understood superconductor, i.e.~Al. We then apply our hybrid resonator technique to measure the kinetic inductance of $2H-$ niobium diselenide (NbSe$_2$), thus demonstrating the effectiveness of our method on an exfoliated TMD superconductor. Along the way, we shall present the theory of kinetic inductance, clarifying the meaning and interpretation of these measurements. Some of the content has been discussed in previous works \cite{Quintana.2014,Antony.2021} but we will include the generic cases and interpret the formulas for practitioners in this growing field.
%Finally, kinetic inductance is an important ingredient in superconducting quantum circuitry: it can be used to engineer qubits\cite{Winkel.2020}, to realize high-sensitivity photon detectors\cite{Day.2003}, and to parametrically amplify signals with performance constrained solely by quantum noise\cite{Eom.2012,Ranzani.2018}. 
% NLME sentences?
% Measurements of these quantities can provide valuable insights into the structure of the superconducting gap, and provide deeper knowledge of the interplay between amplitude and phase fluctuations \cite{Singh.2018}. 

\section{Theory of Kinetic Inductance and Superfluid Stiffness }
In a superconductor, kinetic inductance arises from the inertial response of Cooper pairs to an applied current \cite{meservey1969measurements}. If all the Cooper pairs contribute to the supercurrent, we can equate the total kinetic energy to kinetic inductance and derive
\begin{equation}
    L_K = \dfrac{m^*}{2 n_s e^2} \left( \dfrac{l}{wt}\right) %\mathcal{D}_s = \frac{1}{8}\frac{\hbar^2}{e^2L_K}\frac{l}{wt} = \frac{1}{4}\frac{\hbar^2n_s}{m^*},
\end{equation} 
where $n_s$ is the superfluid density, $t$ is the thickness of the sample component, and $e$ and $m^*$ are the charge and effective mass per quasiparticle, such that the mass of a Cooper pair is $2m^*$. 
The simple formula is insightful, but assumes Galilean invariance in order to relate superfluid density to kinetic inductance. This argument typically does not apply to real materials.
More broadly, kinetic inductance measures the superfluid stiffness, $\mathcal{D}_s$. An applied current induces a vector potential ${\bf A}({\bf r})$ inside the superconductor that leads to a gradient in the phase $\nabla\phi$ of the order parameter, $\Delta({\bf r}) \rightarrow \Delta({\bf r})\exp({ i~2e {\bf A}({\bf r})\cdot {\bf r}/\hbar})$. The factor of 2 in the phase arises because the Cooper pairs involve two electrons, each of which picks up one unit of phase.
The deformation costs energy $\epsilon$ which is controlled by the stiffness
\begin{equation}
    \epsilon = \int d^3r \mathcal{D}_s(\nabla\phi)^2, \label{eq:EnergyStiffness}
\end{equation}
and hence, kinetic inductance can directly measure the cost of phase fluctuations. However, an exact relation between kinetic inductance and superfluid stiffness depends on details of the device. The subtleties arise from the difficulty of finding ${\bf A}({\bf r})$ inside the superconductor which can depend on sample geometry, London penetration depth, coherence length and mean free path (Appendix \ref{sec:lambda_app} for details). However, if we assume a uniform current profile that affects all Cooper pairs, we can derive the expression
\begin{equation}
    L_K = \dfrac{1}{8} \left( \dfrac{\hbar^2}{e^2 \mathcal{D}_s} \right) \left( \dfrac{l}{wt}\right) \label{eq:stiffness},
\end{equation}
which establishes a direct connection between superfluid stiffness and kinetic inductance.

Apart from the spatial profile of ${\bf A}({\bf r})$, we have assumed both the locality and isotropy of the Mattis-Bardeen kernel \cite{Mattis.1958} to relate the current density to the vector potential as explained in the Appendix \ref{sec:lambda_app}: 
\begin{equation}
    J_\mu({\bf r}) \propto \mathcal{D}_s \delta_{\mu\nu} {\bf A}_\nu({\bf r}),
\end{equation}
where $\mu,\nu$ denote spatial directions. While the more general Mattis-Bardeen kernel offers a complete description, it complicates the analysis and obscures the relationship between kinetic inductance, penetration depth, and superfluid stiffness. Specifically, the superfluid stiffness is defined as the ${\bf q}=0$ component of the Fourier transform of the Mattis-Bardeen kernel, $K({\bf q})$ \cite{Tinkham}. The penetration depth is derived from the solution of the non-local London equation \cite{Tinkham}, whereas the kinetic inductance characterizes an effective penetration depth that depends on both $K({\bf q})$ and the boundary conditions \cite{GaoThesis}. Next, we address each assumption individually.

As we discuss in Appendix~\ref{sec:lambda_app}, while the locality assumption might seem invalid for superconductors with large coherence lengths, practical factors such as sample thickness and electronic mean free path strongly reinforce its validity. One may heuristically think of an effective coherence length that is dominated by the smallest of the electronic mean free path and the sample thickness. 
As a result, despite the potential complications of a non-local kernel, the local approximation retains widespread applicability, particularly in thin samples. 

In contrast, the breakdown of isotropic response in anisotropic superconductors is unavoidable. In these systems, the kinetic inductance depends on the alignment of current flow with the crystalline axes of the material. Beyond geometric factors related to the effective penetration volume (as detailed in Appendix~\ref{sec:lambda_app}), the kinetic inductance picks up the penetration depth along the direction of the current.

\section{Description of the measurement method and circuit modeling}
\begin{figure}[t]
\includegraphics[width=0.9\columnwidth]{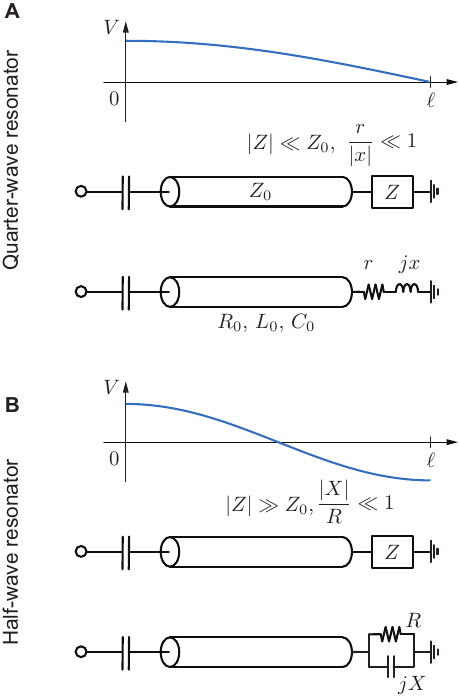}
\caption{Circuit model of (\textbf{A}) quarter-wave resonator and \textbf{(B)} half-wave resonator. The blue curve is a sketch of the magnitude of the voltage oscillations versus the distance from the coupling capacitor. $\ell$ is the physical length of the transmission line, schematically represented by the coaxial cable. The inductor or capacitor of interest with an impedance $Z$ is embedded at the end of the transmission-line resonator to create a quarter-wave, for $|Z| \ll Z_0$, and half-wave, for $|Z| \gg Z_0$, resonator. Without loss of generality, $Z=r+jx = (1/R+1/jX)^{-1}$.}
\label{fig:schematic}\end{figure}

Fig.~\ref{fig:schematic} shows a schematic representation of the transmission line resonator used to probe the sample component. The resonator is drawn as a coaxial cable of length $\ell$ and characteristic impedance $Z_{0}$. In practice, the resonator may consist of any transverse electromagnetic (TEM) or quasi-TEM transmission line, such as a microstrip or coplanar waveguide. Typically operating at radio and microwave frequencies, the resonator is coupled to the external circuitry via the capacitor on the left-hand side of the diagram. The sample with impedance $Z$ terminates the resonator. We model the sample as an equivalent circuit with $Z=r+jx$, where $j = \sqrt{-1}$. The reactive component $x$ modifies the resonance frequency, while the dissipative  component $r$ degrades the $Q$ factor of the probe resonator by providing an additional channel for energy loss.  
The $r$ accounts for loss in the sample and is agnostic to the origin of the dissipation, which can be due to the electrical contact, dielectric loss, or quasiparticles in the superconductors. The loss tangent, $\tan{\delta}$, of the sample component is defined to be $r/|x|$

Below we analyze the circuit response using perturbation theory in two regimes. We first consider the case where $|Z| \ll Z_0$, which is the case relevant for measurements of KI in superconducting samples. In such a case, the sample effectively shorts the resonator to ground, forming a quarter-wave resonator, and $Z=r+jx$, with $x=\omega L$ ( Fig.~\ref{fig:schematic}A). For completeness, we also consider the case wheere $|Z| \gg Z_0$, which will be of use for probing capacitive components. In this case, the sample effectively creates an open circuit, forming a half-wave resonator, and  $1/Z = 1/R+1/jX$, with $X=1/\omega C$ (Fig.~\ref{fig:schematic}B). In both cases, the quarter-wave and half-wave transmission-line resonators are equivalent to a lump parallel or series RLC resonator near their resonance frequencies, respectively \cite{Pozar.2005}, with $R_0$, $L_0$, and $C_0$ as the resistance, inductance, and capacitance of the probe resonators listed in Table \ref{tab:RLC} in Appendix~\ref{sec:DeriveAppendix}.

\begin{table*}[t]
\centering
\begin{tabular}{ l c c c c} 
\hline%\noalign{\hrule height 0.8pt}%\hline
&~~~~~~~~~~~~& Quarter-wave resonator &~~~~~~~~~~~~& Half-wave resonator \\
\hline%\noalign{\hrule height 0.8pt}
Resonant-frequency shift, $\delta\omega_0$ & & \parbox{2.5cm}{\begin{equation} \delta\omega_0=-\frac{2}{\pi}\frac{x}{Z_0}\omega_0\nonumber\end{equation}} & & \parbox{2.5cm}{\begin{equation} \delta\omega_0=\frac{1}{\pi}\frac{Z_0}{x}\omega_0\nonumber\end{equation}}\\ 
\hline
Quality factor, $Q_i$ & & \parbox{4cm}{\begin{eqnarray}
\frac{1}{Q_i} &=& \frac{1}{Q_0}+\frac{1}{Q_{\rm eff}}\nonumber\\
&=& \frac{4}{\pi}Z_0\left(\frac{1}{R_0}+\frac{1}{Z_0^2/r}\right)\nonumber\\
&=& \frac{1}{Q_0}+\frac{4}{\pi}\frac{|x|}{Z_0}\tan{\delta}\nonumber\\ &=&\frac{1}{Q_0}+2\left|\frac{\delta\omega_0}{\omega_0}\right|\tan{\delta}\nonumber\end{eqnarray}} & & \parbox{4cm}{\begin{eqnarray}
\frac{1}{Q_i} &=& \frac{1}{Q_0}+\frac{1}{Q_{\rm eff}}\nonumber\\
&=& \frac{2}{\pi}Z_0\left(\frac{1}{R_0}+\frac{1}{R}\right)\nonumber\\
&=& \frac{1}{Q_0}+\frac{2}{\pi}\frac{Z_0}{|x|}\tan{\delta}\nonumber\\ &=&\frac{1}{Q_0}+2\left|\frac{\delta\omega_0}{\omega_0}\right|\tan{\delta}\nonumber\end{eqnarray}}\\ 
\hline
\end{tabular}
\caption{\textbf{Formulae for quality factor and resonant-frequency shift.}}
\label{tab:formulae}
\end{table*}

\subsection{Quarter-wave resonator circuit model}
Consider a quarter-wave transmission-line resonator of length $\ell$ connected to the ground plane without the sample component. The input impedance of the resonator can be computed \cite{Pozar.2005} by means of the Telegrapher equations, and is equivalent to that of a parallel RLC circuit with resonance frequency $\omega_0=\pi v/2\ell$, where $v$ is the velocity of light in the transmission line. The quality factor of the superconducting probe resonator, $Q_0$, is given by:
\begin{eqnarray}
Q_0 &=& \omega_0R_0C_0 =\frac{R_0}{\omega_0L_0} = \frac{\pi}{4}\frac{1}{\alpha\ell}\label{eq:QWave0Q}\end{eqnarray} where $\alpha$ is the microwave attenuation per unit length of the transmission line. The microwave loss, which can be due to quasiparticles or two-level systems \cite{Sage.2011,Gruenhaupt.2018}, is accounted for using the shunt resistor $R_0$ in the parallel $RLC$ resonator model. A larger $R_0$ results in less electrical current going through the dissipative resistor and thus, a higher $Q_0$. 

Using the Telegrapher equation, the input impedance of the resonator now becomes:
\begin{eqnarray}
Z_{in} &=& Z_0 \frac{(r+jx)+Z_0\tanh{(\alpha+jk)\ell}}{Z_0+(r+jx)\tanh{(\alpha+jk)\ell}}\nonumber\\
&\simeq& \frac{Z_0}{\alpha\ell+r/Z_0-j\cot{k\ell}+jx/Z_0}\label{eq:QWave}\end{eqnarray} where $k$ is the wavenumber. The approximation in Eqn.~\ref{eq:QWave} uses $|r+jx| \ll Z_0$, so that the sample component introduces only a small perturbation to the quarter-wave resonator impedance. By comparing the transmission line resonator with the parallel RLC resonator and relating their physical parameters (see Appendix~\ref{sec:DeriveAppendix}), the loss becomes $\alpha\ell + r/Z_0$ with a frequency shift, $\delta\omega_0$, given by:
\begin{equation}
    \delta\omega_0 \simeq -\frac{2}{\pi}\frac{x}{Z_0}\omega_0.\label{eq:Qshift}
\end{equation}
The resonance-frequency shift is negative for an inductive sample, which effectively elongates the electrical length of the resonator.

The new internal quality factor $Q_i$, for the quarter-wave resonator with the sample component becomes:
\begin{eqnarray}
\frac{1}{Q_i} &=& \frac{1}{Q_0}+\frac{1}{Q_{\rm eff}}\label{eq:TotalQwaveQ}\\
\text{where}~~\frac{1}{Q_{\rm eff}} &=& \frac{4}{\pi}\frac{r}{Z_0}\nonumber\\
&=& 2\left|\frac{\delta\omega_0}{\omega_0}\right|\tan{\delta}\label{eq:QwaveQ}\end{eqnarray}
This $Q_{\rm eff}$  term has two contributions: the loss tangent of the sample component $\tan{\delta}$, and the participation ratio $|\delta\omega_0/\omega_0|$, which weights it. The participation ratio is the ratio of  energy stored in the sample component with respect to the total stored energy in the resonator \cite{Minev.2021}. A high $Q_0$ is necessary to reliably extract the loss of the sample component from the measurement of the $Q_i$. Using Eqn.~\ref{eq:QWave0Q} and \ref{eq:QwaveQ}, we can rewrite Eqn.~\ref{eq:TotalQwaveQ} as
\begin{equation}
    \frac{1}{Q_i} = \frac{4}{\pi}Z_0\left(\frac{1}{R_0}+\frac{1}{Z_0^2/r}\right).
    \label{eq:Q_Quarterwave}
\end{equation} The loss from the reactive element effectively acts as a shunt resistor, of resistance $Z_0^2/r$, to the loss of the resonator, i.e. $R_0$.

\subsection{Half-wave resonator circuit model}
Similarly, we can compute the frequency shift and $Q$ factor of the half-wave resonator terminated by the sample component. The $Q$ factor of the bare, half-wave resonator is given by:
\begin{eqnarray}
Q_0 &=& \omega_0R_0C_0 =\frac{\omega_0L_0}{R_0} = \frac{\pi}{2}\frac{1}{\alpha\ell}\label{eq:Hwave0R0}\end{eqnarray} 
 As before, we assume that the sample causes a small perturbation to the transmission line resonator. For a capacitive sample with small loss tangent, we can further approximate that  $R\simeq x/\tan{\delta}$ and $X\simeq x$. The input impedance of the half-wave resonator then becomes:
\begin{eqnarray}
Z_{in} &=& Z_0 \frac{(1/R+1/jX)^{-1}+Z_0\tanh{(\alpha+jk)\ell}}{Z_0+(1/R+1/jX)^{-1}\tanh{(\alpha+jk)\ell}}\nonumber\\
&\simeq& \frac{Z_0}{\alpha\ell+Z_0/R+j\tan{k\ell}-jZ_0/X}\label{eq:HWave}\end{eqnarray}
Comparing with the half-wave transmission line equation \cite{Pozar.2005}, the loss becomes $\alpha\ell + Z_0/R$ with a frequency shift of:
\begin{equation}
    \delta\omega_0 \simeq \frac{1}{\pi}\frac{Z_0}{X}\omega_0,
\end{equation} i.e.~$\delta\omega_0$ is positive (negative) if the element is inductive (capacitive), effectively shortening (elongating) the electrical length of the resonator---opposite the sign of the quarter-wave result (Eqn.~\ref{eq:Qshift}).

The new $Q_i$ of the half-wave resonator becomes:
\begin{eqnarray}
\frac{1}{Q_i} &=& \frac{1}{Q_0}+\frac{1}{Q_{\rm eff}}\label{eq:TotalHwaveQ}\\\text{where }\space
\frac{1}{Q_{\rm eff}} &=& \frac{2}{\pi}\frac{Z_0}{R}\nonumber\\
&=& 2\left|\frac{\delta\omega_0}{\omega_0}\right|\tan{\delta}.\label{eq:HwaveNewQ1}\end{eqnarray} This $Q_{\rm eff}$  term has two contributions: the loss tangent of the sample component $\tan{\delta}$, and the participation ratio $|\delta\omega_0/\omega_0|$, which weights it. Moreover, the loss of the reactive element in a half-wave resonator $r$ effectively acts as a shunt resistor $R$ to that of resonator, i.e. $R_0$.
Using Eqns.~\ref{eq:Hwave0R0} and \ref{eq:TotalHwaveQ}, we can rewrite Eqn.~\ref{eq:HwaveNewQ1} as: \begin{equation}
    \frac{1}{Q_i} = \frac{2}{\pi}Z_0\left(\frac{1}{R_0}+\frac{1}{R}\right)
\end{equation}

\section{Validation of circuit model and experimental technique using BCS theory}
We validate the quarter-wave model developed above by measuring the kinetic inductance of a thin strip of Al, a well-known BCS superconductor, and then demonstrate the utility of this methodology by applying it to the exfoliated TMD superconductor NbSe$_2$. We report our measurements from five sample components: two Al strips and three exfoliated NbSe$_2$ strips. Their dimensions are listed in Table~\ref{tab:SampleDimensions}.

\begin{table}[h]
\begin{tabular}{ l c c c } 
\hline
sample~~~~~&~~~~width~($\mu$m)~~~~&~~~~length ($\mu$m)~~~~&~~~~thickness (nm)~~~~\\
\hline
\multicolumn{4}{c}{Al sample components}\\
\hline
\textbf{KI06} &  $10$  & $420$ & $80$ \\
KI07 &  $10$  & $420$ & $80$ \\
\hline
\multicolumn{4}{c}{NbSe$_2$ sample components}\\
\hline
KI03 &  $25 \pm 3$  & $37.7 \pm 0.1$ & $55.3 \pm 0.4$ \\
KI19 & $8.2 \pm 0.6$ & $26.3 \pm 0.2$ & $36.0 \pm 0.3$ \\
\textbf{KI20} & $5.5$ & $19.0$ & $49 \pm 0.5$ \\
\hline
\end{tabular}
\caption{\textbf{Dimensions of sample components.} Sample components presented in the main text are indicated by bold typeface.}
\label{tab:SampleDimensions}
\end{table}

\subsection{Kinetic inductance of Al strip}
The coplanar waveguide resonators in our experiments are fabricated using procedures that are conducive to measuring exfoliated flakes of superconductors while maintaining high $Q$ factor \cite{Antony.2021}. The process begins with the fabrication of resonators by photo-lithography over a 240~nm sputtered-Nb film of about 2~k$\Omega$ sheet resistance on a float-zone silicon wafer. We use CF$_4$ reactive-ion to etch through the Nb film, followed by an oxygen etch to remove any NbF, and lastly, an Ar etch to remove unwanted surface oxides. Holes are periodically etched into the Nb ground plane to trap stray flux vortices; oxygen plasma is used after photolithography to remove residual photoresist, which could otherwise cause Si pillars to form within the coplanar waveguide; buffered-oxide etch with hydrofluoric acid is used following the etching of the Si trenches to eliminate any remaining NbF resulting from CF$_4$ ion etching. Once the wafer is diced, electron-beam lithography and evaporation is used to deposit an 80-nm-thick Al strip, our sample component, onto an etched Si window at the end of the resonator.

\begin{figure*}[t]
\includegraphics[width=1.8\columnwidth]{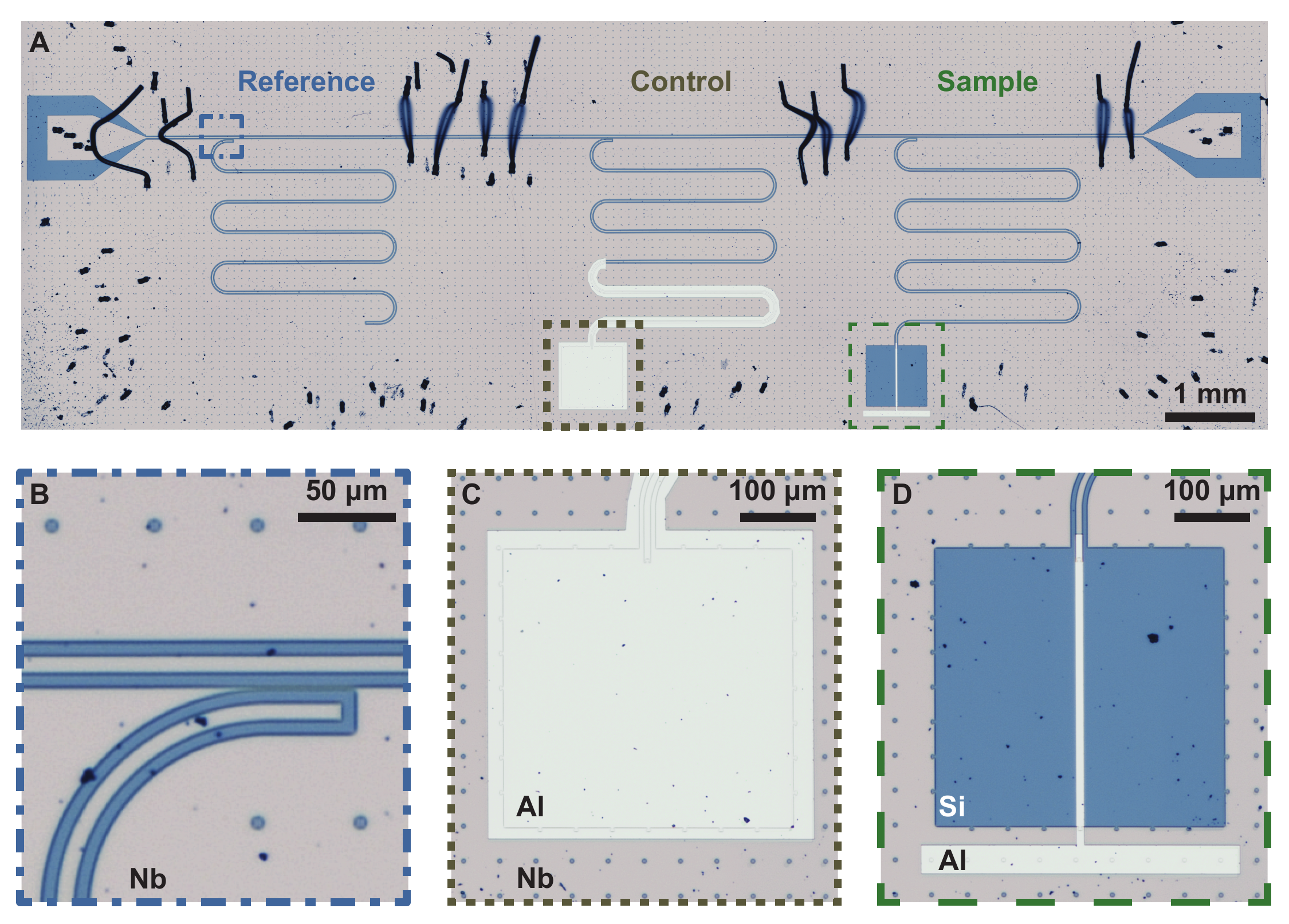}
\caption{\textbf{(A)} Optical image of three Nb coplanar waveguide resonators (`reference', `control', and `sample') capacitively coupled to a transmission feedline on a Si wafer surrounded by electromagnetic shielding within a dilution refrigerator. Wire bonds (black) connect the Nb ground plane above and below the feedline. \textbf{(B)} 100x magnification optical image of Nb reference resonator capacitively coupled to transmission feedline. \textbf{(C)} Quarter-wave control resonator with 80~nm of evaporated Al terminating the resonator to the ground plane. \textbf(A-D) Si hole arrays are used to pin vortices that may form in the superconducting Nb ground plane. \textbf{(D)} Quarter-wave sample resonator with 80-nm-thick sample strip of Al evaporated onto a Si window etched into the Nb, and terminated to the Nb ground plane.}\label{fig:Alresonators}\end{figure*}

Fig.~\ref{fig:Alresonators} displays the three coplanar waveguide resonators (from left to right): `reference', `control', and `sample' resonators. The three high-$Q$ ($Q\sim10^{5}$) resonators of prescribed lengths are capacitively coupled to a transmission feedline. The length of each resonator is designed to yield a resonance frequency in the range of 3--9~GHz, to match the cryogenic amplifier and circulator bandwidth. The choice of different resonance frequencies  allows for measurement of multiple sample components per chip for each experimental cool-down. We also maximize the separation between reference and test resonator frequencies to help identify the correct resonator during measurement. When we are uncertain in identifying a resonance, we can temporarily eliminate the coplanar waveguide to the ground plane with a drop of silver epoxy, and recover it by dissolving the epoxy in an acetone bath.

The reference resonator on the far left in Fig.~\ref{fig:Alresonators}A is a half-wave resonator consisting only of Nb, with no additional material. Fig.~\ref{fig:Alresonators}B displays the capacitive coupling between the resonators and the transmission line, which sets the resonators' coupling $Q$ factor $Q_c$. By measuring the resonance frequency and $Q$ factor of the reference resonators as a function of temperature, we quantify any background kinetic inductance and dissipation from the Nb resonators. Panels C and D show the Al patterns of the sample component and control resonators, respectively. The control resonator is terminated directly to the ground plane without the long Al sample component that we intend to measure. The role of the control resonator is to monitor the temperature dependence of the frequency and quality factor of the niobium resonator used to measure our sample.

The resonators are cooled in a dilution refrigerator, inside a microwave package that is optimized to avoid undesired resonance modes near the designed resonance frequencies. The probe signal is sent through a highly attenuated coax cable to eliminate the thermal noise from room temperature. The signal then passes through the transmission bus line into the resonators where it interrogates the sample before proceeding to the cryogenic low-noise amplifier through a circulator (see Appendix~\ref{sec:SI} for setup schematics). We measure the resonances from the scattering parameter, $S_{21}$, which is defined as the ratio of the output to input voltage waves of the transmitted microwave signal through the feedline.

\begin{figure}[t]
\includegraphics[width=1\columnwidth]{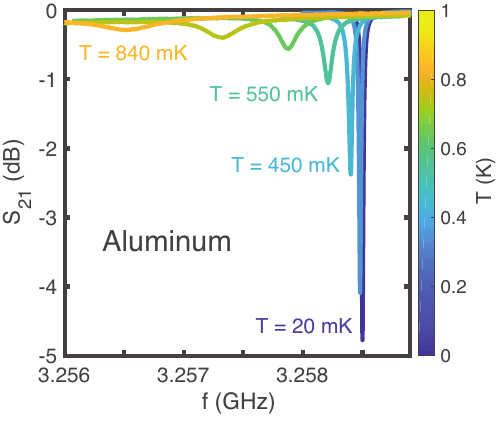} 
\caption{Temperature-dependent $S_{21}$ of the quarter-wave resonator containing an inductive strip of Al measured using a power of -50~dBm from the microwave source. Resonance frequency and loaded $Q$ factor decrease with increasing temperature due to the increasing number of thermally excited quasiparticles.}\label{fig:S}\end{figure}

Fig.~\ref{fig:S} plots $S_{21}$ of the sample component resonator over a range of temperatures from 20 to 840~mK. At the resonance frequency, the signal couples into the quarter-wave resonator, resulting in sharp dip in $S_{21}$. We measured $S_{21}$ for a range of probe powers to ensure that the power is low enough and would cause no shift in the resonance frequency due to heating from the input power or non-linearity from the Kerr effect. The resonance is sharpest at the lowest temperature, and shows a gradual decrease in resonance frequency and $Q$ factor as the temperature increases, due to the increase in kinetic inductance and dissipation, respectively.

\begin{figure}[t]
\includegraphics[width=1\columnwidth]{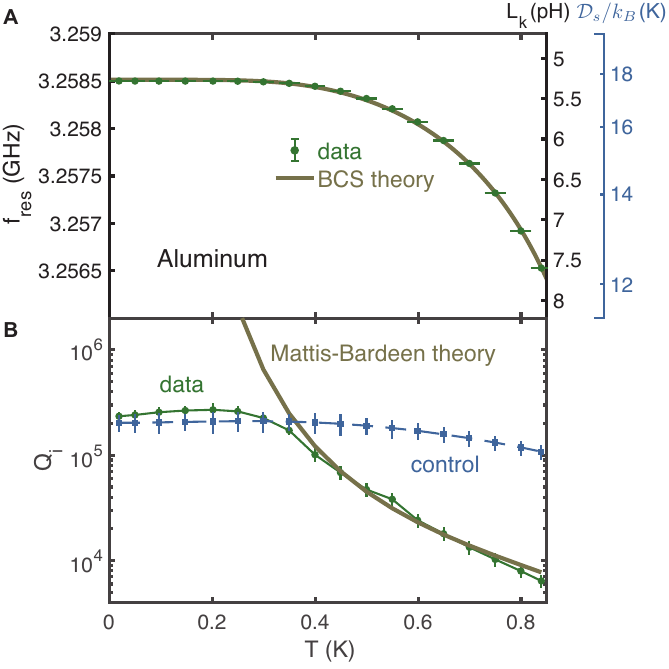}
\caption{\textbf{(A)} Resonance frequency (green data points) and \textbf{(B)} internal-$Q$ factor $Q_i$ of the quarter-wave resonator containing an Al sample component. (A) Solid brown curve is the best fit to BCS theory, i.e. Eqn.~\ref{eqn:FreqShiftData} and \ref{eq:BCS_KI}. Cooper pair density in Al decreases with increasing temperature, thus increasing kinetic inductance. Right y-axis portrays the data as kinetic inductance in units of pH and $\mathcal{D}_s$ in units of K. (B) Using the fitted parameters in (A), we calculate $Q_i$ according to the Mattis-Bardeen theory of surface impedance (solid brown line). The remarkable agreement to the data suggests the degradation of $Q_i$ for $T \geq 0.4$~K is dominated by thermally excited quasiparticles in the sample component rather than the resonator because the corresponding drop in $Q_i$ for the control resonator in the same temperature range is much smaller (blue dashed line). For $T < 0.4$~K, however, $Q_i$ is potentially dominated by the resonator, not the Al strip.}\label{fig:AlfnQ}\end{figure}

To eliminate the effect of the microwave environment, including interference due to multiple reflections of the probe signal along the measurement chain, we analyze the $S_{21}$ data using the circle fitting method \cite{Probst.2015} to extract the resonance frequency as well as the internal quality factor $Q_i$ of the microwave resonator, which are plotted in Fig.~\ref{fig:AlfnQ}A and B, respectively. When applying the equation for the frequency shift of the quarter-wave resonator in Table~\ref{tab:formulae}, we use the definition of $\Delta f = f(T) - f_0$ and $x = 2\pi f_0L_K(T)$ where $f(T)$, $f_0$, and $L_K(T)$ are the resonance frequency at zero temperature, i.e. $T=0$ K, and the kinetic inductance of the Al, respectively. Using Eqn.~\ref{eq:Qshift}, we have
\begin{equation} f(T)  = f_0\left(1-4\frac{f_0L_K(T)}{Z_0}\right). \label{eqn:FreqShiftData}\end{equation}
To fit Eqn.~\ref{eqn:FreqShiftData} to our data, we use BCS theory for the kinetic inductance \cite{Annunziata.2010}:
\begin{equation} L_K = \frac{l}{w}\frac{R_\square\hbar}{\pi\Delta}\coth\left(\frac{\Delta}{2k_BT}\right)\label{eq:BCS_KI}\end{equation}
where $\hbar$ is the reduced Planck's constant, $k_B$ the Boltzmann constant, $T_c$ the critical temperature,  $l = 420~\mu$m the length, $w = 10~\mu$m the width, $R_\square$ the sheet resistance, and $\Delta$ the BCS temperature-dependent gap energy of the Al strip. Using the approximation of $\Delta = \Delta_0[1-(T/T_c)^4]^{1/2}$ with $\Delta_0 = 1.764\cdot k_BT_c$, we employ a least-squares fit of the data to Eqn.~\ref{eqn:FreqShiftData} with three parameters: $T_{c}$, $R_{s}$, and $f_{0}$, and plot the result as the brown line in Fig.~\ref{fig:AlfnQ}A. We obtain $f_{0}$, the resonance frequency without the contribution of $L_K(T)$, from the fitting rather than the measured control resonator frequency because the resonance frequencies of lithographically identical resonators may vary by approximately 3~MHz, possibly due to slight fabrication differences or due to interaction between two resonances at the same frequency (see Fig.~\ref{fig:Uncertainty} in Appendix~\ref{sec:SI}). This statistical variance is slightly less than one part per thousand (ppt), but sizable when compared to the shift of resonance frequency induced by the kinetic inductance. This uncertainty in $f_0$ constitutes the limitation in accuracy of this measurement method. However, this measurement method has high precision, inherited from the high-$Q$ factor of the resonators. As such, the error bar of the resonance frequency in Fig.~\ref{fig:AlfnQ}A is less than one part per million (ppm).

\begin{table}[h]
\begin{tabular}{ l c c } 
\hline
%\multicolumn{3}{c}{~~~~~~~~Josephson Junction Device Parameters~~~~~~~~}\\
Quantity~~~~~~~~~&~~~~~~~~Fitted value~~~~~~~~&~~~~~~~~DC transport~~~~~~~~\\
\hline
$f_0$ (GHz) &  $3.263 \pm 0.001$  & $\ldots$ \\
$T_c$ (K) & 1.195 $\pm$ 0.001 & 1.20 $\pm$ 0.01 \\
$R_\square$ (m$\Omega$) & 109 $\pm$ 26 & 75 $\pm$ 20 \\
\hline
\end{tabular}
\caption{\textbf{Fit parameters extracted from frequency data fitted to BCS theory for the Al sample component shown in Fig.~\ref{fig:AlfnQ}A.}}
\label{tab:FitValues}
\end{table}

Table \ref{tab:FitValues} shows the fit parameters extracted from the BCS model. We can compare the fitted values of $T_{c}$ and $R_{s}$ with their respective measured values using DC transport on an Al film that was deposited alongside the resonator sample component. For the Al sample component, fit values are $T_c = 1.195 \pm 0.001$~K and $R_{s} = 109 \pm 26$~m$\Omega$ which both overlap with the measured DC values of $T_{c}$ = 1.20 $\pm$ 0.01~K and $R_\square = 75.5 \pm 19.6$~m$\Omega$. Using the fitted values, we calculate the kinetic inductance of the superconducting Al sample component and plot it on the right axis in Fig.~\ref{fig:AlfnQ}A. Eqn.~\ref{eq:BCS_KI} agrees well with the measured data. The zero-temperature kinetic inductance $L_{k0} \simeq$ 7~pH, nearly 10$^3$ smaller than the equivalent inductance of the quarter-wave resonator, i.e. $\simeq$2.9~nH, confirming that we are well within the perturbation regime and free to apply the equations in Table~\ref{tab:formulae}.

As discussed above, measuring both $\mathcal{D}_s$ and $\Delta$ as a function of carrier density, temperature, magnetic field, and bias current is essential in revealing the true nature of superconductivity \cite{Emery.1995,Singh.2018}. Using Eqn.~\ref{eq:stiffness}, we find that $\mathcal{D}_s/k_B$ (see right $y$-axis of Fig.~\ref{fig:AlfnQ}A) is on the order of a few Kelvins, i.e. $> \Delta_0/k_B$, suggesting that it takes more energy to create phase fluctuations than to break apart Cooper pairs in Al. We can further estimate the London penetration depth using 
\begin{equation}
    \lambda_L^2 = \dfrac{\hbar^2}{4 \mu_0 e^2 \mathcal{D}_s} = 2\frac{w t}{l}\frac{L_K}{\mu_0},\label{eq:London}
\end{equation}
where $\mu_0$ is the vacuum permeability, to be about 126~nm at $T \simeq$~0~K, in agreement with the $\lambda_L$ obtained from similar resonator methods \cite{GaoThesis,Lopez-Nunez.2023}. The applicability of the London equation to Al is somewhat surprising because its kinetic inductance could be in the anomalous limit (see Appendix~\ref{sec:lambda_app}), due to its long coherence length and mean free path. However, as we discussed earlier, the film thickness restricts the size of Cooper pairs and drives the thin film into a quasi-local limit \cite{Lopez-Nunez.2023}.

By measuring the quality factors of the bare resonator and ($Q_0$) and the sample-terminated resonator ($Q_i$), we can infer the quality factor of the sample component ($Q_{\rm{eff}}$). This value then yields the loss tangent $\delta$ (Eqn. \ref{eq:QwaveQ}) at the resonance frequency. The loss tangent $\tan{\delta} =\sigma_1/\sigma_2$ is equivalent to the ratio of the resistive and reactive components of the complex conductivity $\sigma = \sigma_1 - j\sigma_2 $. We note that here we follow the negative sign convention first introduced by Glover and Tinkham \cite{Glover.1957} wherein microwave photons were used to probe the complex conductivity. In fact, Glover and Tinkham reported observation of the superconducting energy gap in far-infrared measurements(submitted May 17, 1957) before the paper describing BCS theory, which formed the basis for the 1972 Nobel prize in physics, was submitted (July 8, 1957). 

Figure ~\ref{fig:AlfnQ}B plots the measured $Q_{\rm{i}}$ of the sample resonator, along with that of a bare control resonator, \textit{vs.} temperature. The solid line shows the expected value of $Q_{\rm{eff}}$, obtained by using the resonance shift from panel A and evaluating the Mattis-Bardeen integrals to obtain $\sigma_1$ and $\sigma_2$ at finite temperatures \cite{Mattis.1958}, where 
\begin{equation}
    \dfrac{1}{Q_{\rm{eff}}} = 2 \dfrac{\delta \omega_0}{\omega_0} \dfrac{\sigma_1}{\sigma_2}.\label{eq:QandSigma}
\end{equation}
For $T \geq 0.4$~K, the measured $Q_{\rm{i}}$ closely follows the calculated $Q_{\rm eff}$, indicating that increasing dissipation with increasing $T$ is driven by a decrease in $\sigma_2/\sigma_1$ due to thermally excited quasiparticles from the breaking of Cooper pairs \cite{Gao.2008}. The close agreement between theory and experiment is remarkable because we use only the fitting parameters from the frequency shift to calculate $\sigma_1$ and $\sigma_2$. For $T < 0.4$ K, on the other hand, $Q_{\rm i}$ is limited by the quality factor of the transmission-line resonator, $Q_0$. 

Examining the data below 0.4~K carefully, we observe that the measured $Q_{\rm i}$ slightly exceeds that of the control resonator and shows a minor peak near 0.2~K. This peak in $Q_i$ is a key signature of the $Q$ factor being limited by dissipation through two-level systems (TLSs). At temperatures well below $\hbar\omega_0/k_B$, the TLSs are in their ground states and ready to absorb photon energy from the measurement source. However, at temperatures well above $\hbar\omega_0/k_B$, the TLSs are already thermally excited out of their ground states, therefore reducing the probability of photon absorption and subsequent losses through the TLS channel. As a result, $Q_i$ peaks at around $\hbar\omega_0/k_B$, i.e. 0.2~K for our sample component resonator, before its eventual degradation due to quasiparticles as temperature increases \cite{Alexander.Richardson.2022}. This minor peak in $Q_i$ is suppressed when measuring the resonator with higher power, because microwave photons from the measurement source can saturate the TLSs. The behavior of $Q_i$ as a function of temperature at various powers, plotted in Fig.~\ref{fig:ThreePower} in Appendix~\ref{sec:SI}, follows this expectation.

\begin{figure}[t]
\includegraphics[width=0.9\columnwidth]{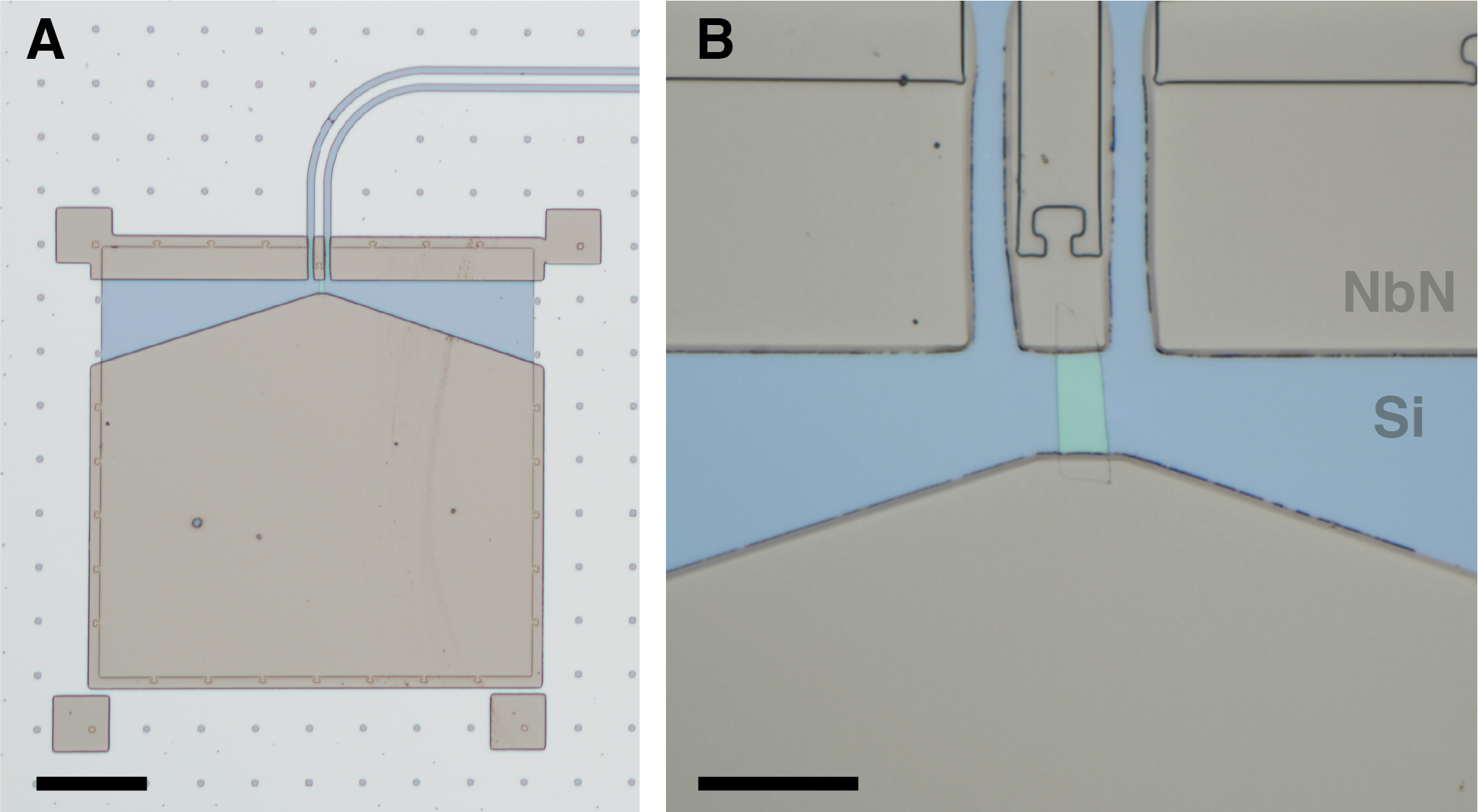}
\caption{Optical image of the niobium diselenide sample embedded in the quarter-wave resonator. \textbf{(A)} Grounding the coplanar waveguide resonator through an exfoliated strip of NbSe$_2$ sample using niobium nitride. The coplanar waveguide and the ground plane are made of Nb (light gray). \textbf{(B)} Close up image of the NbSe$_2$ sample in a light green color. The scale bars in panel (A) and (B) are 100 and 20~$\mu$m, respectively.} %Quarter-wave resonator with 36 nm thick flake of $NbSe_2$ placed onto Si void as inductive sample component. The flake is contacted and terminated to the Nb ground plane with $240$ nm sputtered $NbN$.  \textbf(C) Quarter-wave control resonator with Si void completely covered by $240$ nm of sputtered $NbN$. \textbf(A-D) Silicon holes are periodically spaced throughout the wafer to trap any potential vortices in the superconducting Nb}
\label{fig:NbSe$_2$resonators}\end{figure}

\subsection{Kinetic inductance of a niobium diselenide flake}
We apply the above technique to measure the kinetic inductance of the layered superconductor $2H-$ niobium diselenide (NbSe$_2$). NbSe$_2$ belongs to the family of TMD superconductors and possesses many interesting properties. The Fermi surface consists of a set of pockets at $\Gamma$ and $K$, which exhibit the nesting property \cite{Yokoya.2001, Johannes.2006}, and features an anisotropic superconducting gap \cite{Fletcher.Giannetta.2007,Noat.2015,Khestanova.2018, Dvir.2018}. As the number of layers decreases, the transition temperature becomes gate-tunable and drops from roughly 7.2~K to $< 5$~K \cite{Frindt.1971}, potentially due to the increased role of fluctuations in the 2D limit. Additionally, in the few-layer limit, evidence of nematicity setting in at $T_c$ has been reported, suggestive of a sub-leading $p$- or $d$-wave state which coexists with the $s$-wave state observed in the bulk limit \cite{Hamill.2021}. The locally non-centrosymmetric crystal structure results in strong spin-orbit coupling, believed to result in ``Ising superconductivity'' associated with high critical fields. In experiment, superconductivity is seen to persist up to the application of a $\sim$35 T in-plane magnetic field \cite{Xi.2016}. Other interesting features of this system include the possible observation of a Fulde–Ferrell–Larkin–Ovchinnikov phase \cite{Wan.2023,Liu.Davis.2021}. The kinetic inductance of few-layer-thick NbSe$_2$ gives further insight into the pairing symmetry, and serves as proof of concept of our measurement technique as it applies to exfoliated vdW superconductors. 

Fig.~\ref{fig:NbSe$_2$resonators} displays optical micrographs of an exfoliated NbSe$_2$ strip connected to a transmission line resonator. The fabrication starts with mechanical exfoliation of NbSe$_2$ flakes onto SiO$_2$/Si substrates in an inert N$_2$ environment inside a glovebox with oxygen levels lower than 0.5~ppm. We employ the dry pickup technique \cite{Wang.2013} using polycarbonate stamps to transfer NbSe$_2$ flakes from the SiO$_2$/Si to the destination area on a substrate with pre-patterned Nb resonator structures. After removing polycarbonate residue with chloroform, we pattern the connection between the NbSe$_2$ flake and Nb transmission lines by electron-beam lithography. Next, we perform \emph{in situ} Ar ion milling to remove surface oxide layers from the Nb resonator and NbSe$_2$ flake (Fig.~\ref{fig:NbSe$_2$resonators}B), followed by subsequent deposition of a 3-nm-thick Ti sticking layer and reactive sputtering of 200--240~nm of NbN to make galvanic connection between the two with negligible contact resistance \cite{Antony.2021}. We use a relatively thick NbN not only to ensure that it forms contact over the $\sim$100--300~nm thick NbSe$_2$, but also to reduce the contribution of kinetic inductance from NbN by suppressing its sheet resistance. In addition to the resonator shown in Fig.~\ref{fig:NbSe$_2$resonators}, each NbSe$_2$ device also consists of a Nb reference resonator and an Nb-NbN control resonator (see Fig.~\ref{fig:NbSe2Device} in Appendix). No aluminum is used for the NbSe$_2$ devices.% We determined a milling rate of 10 nm/min for NbSe$_2$ and 6 nm/min for Nb in our deposition tool (see Supplementary Section X).

\begin{figure}[t]
\includegraphics[width=1\columnwidth]{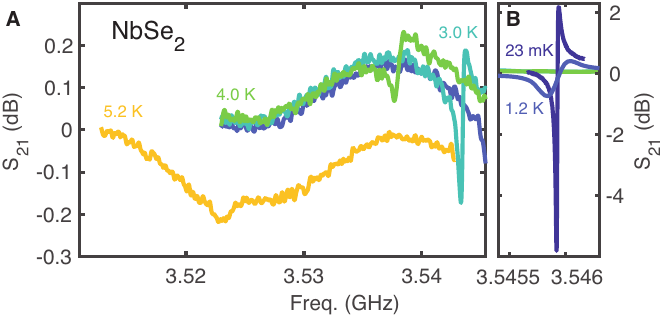}
\caption{Temperature-dependent $S_{21}$ of a hybrid Nb-NbSe$_2$ quarter-wave resonator terminated by a 36~nm thick inductive flake of NbSe$_2$ between the Nb resonator and Nb ground plane. Data taken using a power of -50~dBm from the microwave source, before attenuation in the fridge lines. Resonance frequency and $Q_{\rm i}$ decrease with increasing temperature due to thermally excited quasiparticles.}\label{fig:NbSe$_2$S21}\end{figure}

\begin{figure}[t]
\includegraphics[width=1\columnwidth]{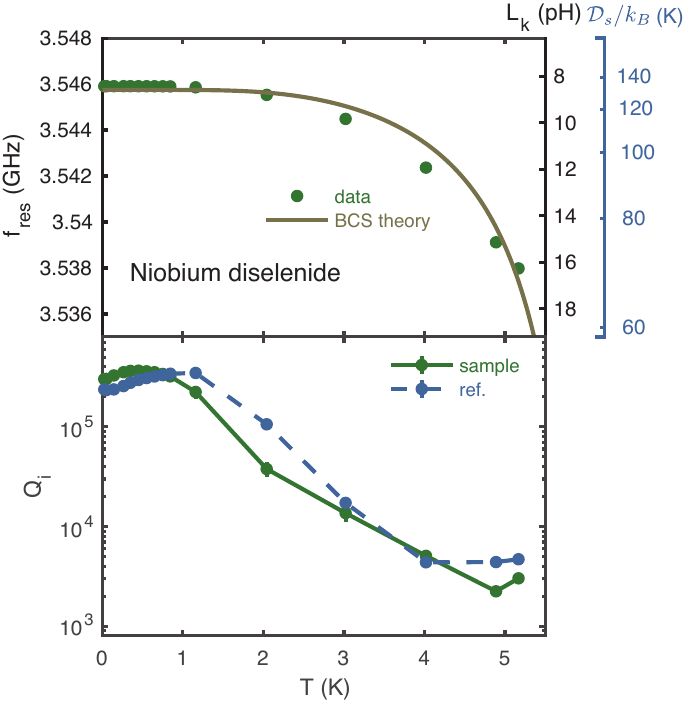}
\caption{\textbf{(A)} Comparison of measurement-extracted (dark green marks) and BCS fit (brown line) kinetic inductance of NbSe$_2$, without contribution from control, as a function of temperature. The dark green measured data is the difference of the normalized frequency shift of the control resonator and the hybrid Nb-NbSe$_2$ sample resonator. Cooper pair density in NbSe$_2$ decreases with increasing temperature, thus increasing kinetic inductance $L_k$ and decreasing superflulid stiffness $D_s$. \textbf{(B)} Semi-log plot of extracted internal quality factor ($Q_i$) as a function of temperature. The green curve corresponds to $Q_i$ of the sample component resonator, whereas the blue dashed curve corresponds to $Q_i$ of the reference resonator (see Fig.~\ref{fig:Alresonators}B}.
\label{fig:NbSe$_2$fnQ}\end{figure}

Fig.~\ref{fig:NbSe$_2$S21} plots the scattering parameter of a hybrid Nb-NbSe$_2$ quarter-wave resonator containing the exfoliated NbSe$_2$ sample component. Similar to the Al-terminated resonator, both the resonance frequency and 
$Q_{\rm i}$ (see Fig.~\ref{fig:NbSe$_2$fnQ}) decrease with increasing temperature. However, the change of the resonance frequency spans a higher temperature range than that shown in Fig.~\ref{fig:AlfnQ} for Al. This is commensurate with the higher transition temperature of NbSe$_2$, 7.06~K, measured by DC electrical transport (see Fig.~\ref{fig:TcNbSe2} in Appendix~\ref{sec:SI}). At these temperatures, we also observe a frequency change for the control and reference resonators. This is expected because both Nb and NbN are superconducting materials with their own kinetic inductance. At temperatures well below their $T_c$ values---measured to be 8.59 and 12.13~K, respectively---their kinetic inductances are negligible in our resonator design. However, at temperatures approaching the $T_c$ of NbSe$_2$, the shift of the NbSe$_2$ resonance frequency includes a contribution from the change in kinetic inductance of the Nb transmission-line resonator.

\begin{figure}[t]
\includegraphics[width=1\columnwidth]{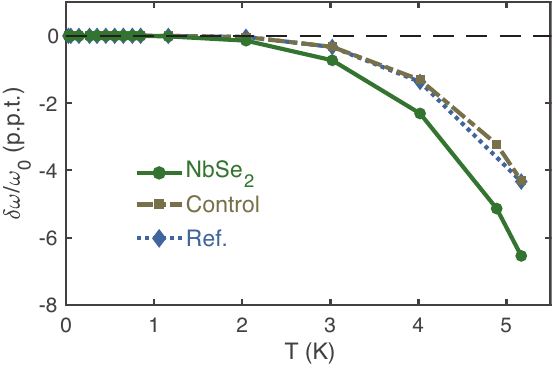}
\caption{Comparison of normalized frequency shift $\delta\omega/\omega_0$ of the three resonators in our NbSe$_2$ experiment. $\delta\omega_0/\omega_0$ of the control and reference resonator are about the same. This is noteworthy because their resonance frequencies are at 3.680 and 8.494~GHz, yet their normalized frequency shifts collapse to the same curve. We later extract the change in kinetic inductance of NbSe$_2$ absent any contribution from the change in kinetic inductance of the transmission-line resonator by taking the difference of their respective $\delta\omega_0/\omega_0$ as shown in Fig.~\ref{fig:NbSe$_2$fnQ}.}
\label{fig:dfOverf_Comparison}\end{figure}

%\textcolor{red}{what is the Ref resonator? I don't see it explained in the text}
% Ref resonator is described in on page 5 in the paragraph beginning "The reference resonator on the far left in Fig. 2A.."
Figure ~\ref{fig:dfOverf_Comparison} plots the fractional frequency shift $\delta \omega/\omega_0$ \emph{vs.} temperature for the NbSe$_2$-terminated, control, and reference resonators. Since $\delta\omega/\omega_0$ follows the fractional change in inductance, we expect this quantity to track the temperature-dependent kinetic inductance in each resonator. Notably, $\delta\omega/\omega_0$ is the same for the control and reference resonators, even though their resonance frequencies are 3.680 and 8.494~GHz. This indicates that the change of their kinetic inductance at these temperatures is mostly due to the Nb. We can then extract $\delta\omega/\omega_0$ due to NbSe$_2$ kinetic inductance by subtracting the value for the control resonator from that of the NbSe$_2$-terminated resonator. Similar methods are used in resonator-based experiments \cite{Cho.Prozorov.2011}.

Fig.~\ref{fig:NbSe$_2$fnQ} plots the frequency of the NbSe$_2$ resonator \emph{vs.} temperature, with the contribution from the transmission line subtracted. The coherence length of NbSe$_2$ is only roughly 10~nm, and therefore sets the shortest length scale between coherence length ($\zeta$) and penetration depth of the NbSe$_2$ sample flake. Consequently, the superconducting NbSe$_2$ is within the local limit ($\zeta \ll \lambda_L$), and we apply Eqn.~\ref{eq:London} in order to infer $\mathcal{D}_s$ and $\lambda_L$. As shown in the right $y$-axis in Fig.~\ref{fig:NbSe$_2$fnQ}A, $\mathcal{D}_s/k_B$ remains larger than the $T_c$ of  NbSe$_2$. At the zero-temperature limit, $\lambda_L \approx$ 440~nm, which is consistent with the reported values \cite{Sonier.Brill.1997, Fletcher.Giannetta.2007}. Any discrepancy between our reported value of penetration depth and previously reported values is probably due to the limitation of using Eqn.~\ref{eq:BCS_KI} to describe the two-gap NbSe$_2$ superconductor, as well as error in the frequency subtraction.% used in Eqn.~\ref{SubtractionEqn}.

\begin{figure}[h]
\includegraphics[width=0.8\columnwidth]{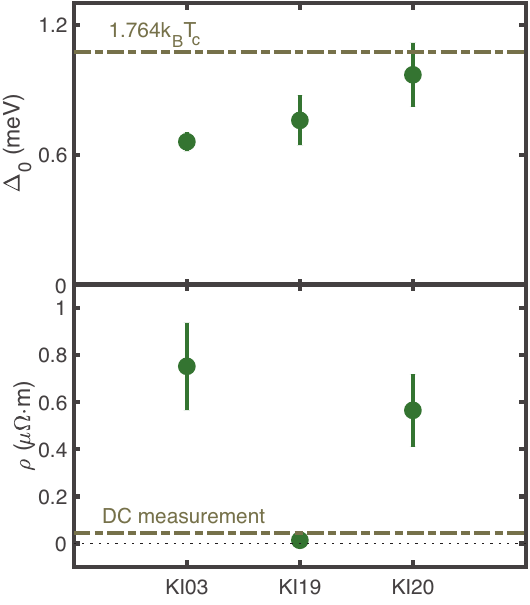}
\caption{Extracted BCS-fit parameter values of \textbf{(A)} zero-temperature gap energy $\Delta_0$ (green points) and \textbf{(B)} resistivity, $\rho$ compared to measured values (dashed lines) from DC transport for three NbSe$_2$ exfoliated strips. Measured gap value in (A) is inferred from measured $T_c = 7.06$ K.}
\label{fig:fittedvalues}\end{figure}

Fig.~\ref{fig:NbSe$_2$fnQ}B plots $Q_{\rm i}$ of both the sample component and reference. Unlike in the temperature range plotted in Fig.~\ref{fig:AlfnQ}B, the $Q_{\rm i}$ of the reference resonators drops considerably because we are now closer to the $T_c$ of Nb. In this temperature range, $Q_{\rm i}$ of the sample component and reference are approximately the same, suggesting that $Q_{\rm i}$ is dominated by the $Q_{\rm 0}$ term rather than the $Q_\text{eff}$ in Eqn.~\ref{eq:QwaveQ}. Since the participation ratio of the sample is merely a few ppt, the probe resonator quality factor $Q_{\rm 0}$ is too low for us to discern the true $Q$ factor of the sample component. This substantiates the limitation of this measurement technique: $Q_{\rm i}$ is only representative of the loss through the sample component when $Q_{\rm 0} \gg Q_{\rm eff}$ in the perturbation regime in which the participation ratio of the sample component is small.

\section{Discussion and conclusion}
Successful application of this new technique requires some careful attention to detail. For example, to suppress the change of kinetic inductance from the superconducting materials used in fabricating the resonator, we sputter 200~nm of Nb to maintain a high geometric-to-kinetic inductance ratio. Nevertheless, as shown in Fig.~\ref{fig:dfOverf_Comparison} when the temperature approaches the $T_c$ of Nb, the frequency shift of the coplanar waveguide resonator can become sizable and complicate the analysis for extracting the reactance and loss of the sample component. Therefore, we monitor both the control and reference resonators (Fig.~\ref{fig:Alresonators}A) to extract the kinetic inductance of the sample component and identify the applicable range of temperatures. We subtract the normalized frequency shift of the control resonator from that of the sample component resonator in order to extract the sample NbSe$_2$ kinetic inductance, but must note that this subtraction may be imperfect due to the inescapable fact that the control resonator is a separate physical object from the sample component resonator. As such, the normalized frequency shift due to the Nb resonator and NbN stitching on the sample component resonator may not be exactly the same as normalized frequency shift of the control resonator, which serves only as a proxy for the actual kinetic inductance contribution from the NbN and Nb of the sample component resonator. Furthermore, one must pay attention to the quality of the electrical contact between the resonator and sample component \cite{Antony.2021}. Extra contact resistance can degrade the $Q$ factor, and thus the measurement sensitivity. If the contact is composed of two different superconductors, the proximity effect may enhance or suppress the superconductivity of the sample component. Using superconductors with a short coherence length compared to the length of the sample component, such as NbN in stitching the NbSe$_2$ to the Nb resonator, can suppress the proximity effect in our measurements. If a Josephson junction were to form at the electrical contact, it would add an extra source of inductance and its critical current would set an additional limitation on the microwave power used in measuring the resonator. For kinetic inductance measurements, we should also consider any nonlocal effects and the relationship between relevant length scales, i.e.~coherence length, mean-free path, and penetration depth.  Fortunately, the typical thickness of exfoliated sample components generally keeps the measurement in the local limit. On the contrary, if the penetration depth is much smaller than the thickness of the sample, more calibration and theoretical calculations would be required to interpret the data from this technique.

In this paper, we present a model for using high-$Q$ transmission-line resonators to measure the inductance or capacitance of a sample component to ppm precision, implement the model by measuring the kinetic inductance of a sample component strip of Al and a sample component flake of NbSe$_2$ using coplanar waveguide resonators, and validate both the model and its implementation by demonstrating agreement between our data and BCS theory for the kinetic inductance and $Q$ factors of the Al and NbSe$_2$ sample components. The reported agreement between data and BCS theory for the kinetic inductance and $Q$ factor vouches for this method as a reliable technique, inviting interrogation of copious vdW materials, as demonstrated by the measurement of NbSe$_2$. Our method is enabled by advances in nanotechnology and quantum sensing and augments traditional techniques for measuring kinetic inductance by the technique of cQED. In addition, our method provides the new capability of measuring both inductance and capacitance, and is now compatible with on-chip DC connections for electrical transport to control the chemical potential of the sample component with a gate, via the field effect \cite{Lee.2020}. Furthermore, we also include the measurement of dissipation, which can be understood from the imaginary and real components of the complex conductivity of a superconductor, and is extremely useful when studying losses in quantum devices \cite{Quintana.2014}. Given its ability to measure mesoscopic flakes of vdW superconductors to ppm precision, this technique will be a highly functional tool in the rise of quantum materials and quantum sensing.

\section{Acknowledgement}
We thank G.~J.~Ribeill for the helpful discussions, and M.~Randeria and T.~Lemberger for their insights and encouragement especially when we just started to have the idea. This work was supported by the STC Center for Integrated Quantum Materials, NSF Grant No. DMR-1231319. K.~C.~F., X.~C., and J.~H.~acknowledge  support from the Army Research Office under Contract W911NF-22-C-0021. 
R.~Q.~acknowledges support from the NSF MRSEC program at Columbia University through the Center for Precision-Assembled Quantum Materials (DMR-2011738).
J.~B.~acknowledges support from the NSF MRSEC program at Columbia University through the Center for Precision-Assembled Quantum Materials (DMR-2011738). The authors acknowledge the use of facilities and instrumentation supported by NSF through the Columbia University, Columbia Nano Initiative, and the Materials Research Science and Engineering Center DMR-2011738.

\section{Appendix}\label{sec:appendix}

\subsection{Theory of kinetic inductance and its relation to the penetration depth}\label{sec:lambda_app}

In this section we establish the theory behind the measurement of kinetic inductance. The kinetic inductance of a superconductor originates from the inertial response of Cooper pairs to the applied current, meaning that the energy of the system $\epsilon = (1/2) L_K I^2$ must match the kinetic energy carried by the condensate. When a current is applied, a vector potential ${\bf A}({\bf r})$ is induced within the superconductor, resulting in an electromagnetic energy.
\begin{equation}
    \epsilon_{\rm EM} = \int d{\bf r}\; {\bf j}({\bf r})\cdot {\bf A}({\bf r}) = \int d{\bf r}\; j_\mu({\bf r}) A_\mu({\bf r})
\end{equation}
where ${\bf j}({\bf r})$ is the current density and $\mu = x,y,z$ is a spatial index with an implied summation over repeated indices. The current that develops in response to the applied vector potential is related in linear response via
\begin{equation}
    j_\mu({\bf r}) = \int d{\bf r}^\prime \; K_{\mu\nu}( {\bf r}- {\bf r}^\prime) A_\nu({\bf r}^\prime).\label{eq:j-a-relationship}
\end{equation}
where $K_{\mu\nu}( {\bf r}- {\bf r}^\prime)$ is the Mattis-Bardeen kernel \cite{Mattis.1958}; this object is otherwise known as the current-current correlator, and can be computed via standard means using the Kubo formula. The response is generically non-local -- i.e., the current at position $\bf r$ depends on the vector potential at $\bf{r} \neq \bf{r}'$. The origin of non-locality is that the response to an electromagnetic field of a superconductor is spread over a region of size determined by the coherence length $\xi$.
In terms of Fourier components, this non-locality brings in a variation in $K({\bf q})$ at the scale of $\xi$.

By matching the electromagnetic energy stored to that of the inductor, we get a general formula of the kinetic inductance
\begin{equation}
    L_K = \dfrac{1}{2I^2} \int d{\bf r} d{\bf r}^\prime \; K_{\mu\nu}( {\bf r}- {\bf r}^\prime) A_\mu({\bf r}) A_\nu({\bf r}^\prime)
\end{equation}
in terms of vector potential inside the superconductor ${\bf A}({\bf r})$ and Mattis-Bardeen kernel $K_{\mu\nu}({\bf r}-{\bf r}^\prime)$. 
As mentioned, there are standard methods for computing $K_{\mu\nu}$ in BCS theory, but computing the vector potential $A$ inside the superconductor requires solving a boundary value problem.

The expression is much simpler if the Mattis-Bardeen kernel is assumed to be local and isotropic 
\begin{equation}
 K_{\mu\nu}({\bf r}-{\bf r}^\prime) = \dfrac{4e^2}{\hbar^2} \mathcal{D}_s \delta({\bf r}-{\bf r}^\prime) \delta_{\mu\nu} \label{eq:approx:local-isotropic-MB-kernel}  
\end{equation}
where $\mathcal{D}_s$ is the superfluid stiffness. This quantity is calculated by computing the momentum dependent quantity $K_{\mu\nu}(\bf q)$ and replacing it with its $q\rightarrow 0$ limit. This expression is valid in the limit where the coherence length $\xi$ is much smaller than the London penetration depth; when $\xi$ is large, the momentum dependence of $K_{\mu\nu}(\bf q)$ can no longer be neglected. As for the vector potential, we can revert Eqn.~\ref{eq:j-a-relationship} to write the kinetic inductance in terms of current density and then make the substitution
\begin{equation}
    \int d{\bf r} j({\bf r})^2 \rightarrow I^2 \left(\dfrac{l^2}{V_{\rm eff}} \right)
\end{equation}
where we have introduced $V_{\rm eff}$ as the penetration volume. It represents the volume of the sample where the current density is finite. Putting it all together, we get
\begin{equation}
    L_K = \dfrac{1}{8} \dfrac{\hbar^2}{e^2 \mathcal{D}_s} \left(\dfrac{l^2}{V_{\rm eff}} \right) = \dfrac{\mu_0 \lambda_L^2}{2} \left(\dfrac{l^2}{V_{\rm eff}} \right).
\end{equation}
If the current density is present throughout the sample, which is the case for thin films ($t\ll \lambda_L$), the penetration volume is equal to the total volume $(=lwt)$ and we recover Eqn.~\ref{eq:stiffness}. On the other hand, if the sample is too thick, current only flows close to the surface on a thin shell of size $\lambda_L$. This reduces the penetration volume to $2lw\lambda_L$ and the resulting kinetic inductance is given by $L_K = (l/w) \mu_0 \lambda_L$.
We can further solve the boundary value problem for arbitary values of $t/\lambda_L$ and interpolate between the two limits with
\begin{equation}
    L_K = \dfrac{\mu_0 \lambda_L}{4} \left(\dfrac{l}{ w}\right) \coth\left( \dfrac{t}{2\lambda_L}\right) .
\end{equation}

The idea of penetration volume is extremely useful in understanding anisotropic responses \cite{Prozorov.2000}.
Anisotropic superconductors have a direction dependent superfluid stiffness, that is, Eqn.~\ref{eq:approx:local-isotropic-MB-kernel} is modified to
\begin{equation}
    K_{\mu\nu}({\bf r}-{\bf r}^\prime) = \dfrac{4e^2}{\hbar^2} [\mathcal{D}_s]_{\mu} \delta({\bf r}-{\bf r}^\prime) \delta_{\mu\nu}
\end{equation}
which leads to a direction dependent penetration depth $\lambda_{L, \mu}$ as well.
Further if $\mu$ denotes the direction of applied current, the element $[\mathcal{D}_s]_{\mu}$ is needed to relate the current density to the vector potential.
In the limit of thin samples such that the penetration volume is the same as the volume of the device, the kinetic inductance then becomes
\begin{equation}
L_{K,\mu} = \left( \dfrac{l}{wt} \right) \dfrac{\mu_0 \lambda_{L,\mu}^2}{2}
\end{equation}
where the only difference from the previous expression is the additional index $\mu$.
More complications arise if the sample is thick such that fields do not penetrate entirely. The penetration volume for a thick slab is then given by
\begin{equation}
    V_{\rm eff} = 2 (lw \lambda_c + lt \lambda_b + wt \lambda_a)
\end{equation}
where in general $\lambda_a \neq \lambda_b \neq \lambda_c$. The kinetic inductance thus has a complicated dependence on the geometry of the device, including not only the dimensions but also aspect ratios \cite{Prozorov.2000, Fletcher.Giannetta.2007}.

Barring details of the device geometry, the local limit always connects the kinetic inductance to the penetration depth $\lambda_L$ and equivalently the superfluid stiffness $\mathcal{D}_s$ as the two are related $1/\lambda_L^2 = (4\mu_0 e^2/\hbar^2) \mathcal{D}_s$. This relation breaks down in the nonlocal regime where the coherence length is larger than the penetration depth --in this case kinetic inductance measures an effective penetration depth which depends on  $K(\bf q)$ \cite{GaoThesis}. 
In this case there are no simple analytic connections between effective penetration depth and London penetration depth or superfluid stiffness. 

It is worth mentioning that all our measurements fall in the local limit. In particular, electronic mean-free path $l_{\rm mfp}$ restricts the size of Cooper pair in real space -- heuristically, the mean free path acts as an upper bound on the effective coherence length. Phenomenologically, it exponentially damps the Mattis-Bardeen kernel \cite{Seibold2017, Mattis.1958} $K(r) \rightarrow K(r) e^{ - |r|/l_{\rm mfp}}$ and hence, short mean-free path can drive the system to a local limit, even if the BCS estimate of coherence length is larger than penetration depth. 
Similar arguments can be made for thin samples where the thickness controls the pair size \cite{GaoThesis}.

\subsection{Derivations of frequency shift and $Q$ factor of resonance}\label{sec:DeriveAppendix}
\subsubsection{Quarter-wave resonator}
For the quarter-wave transmission-line resonator with the end connected directly to the ground plane without going through the sample component, the input impedance of the resonator is given by Telegrapher equation \cite{Pozar.2005}:
\begin{equation}
Z_{in} = Z_0\tanh{(\alpha+j\beta)\ell}\label{eq:QWave0Tele}\end{equation}
For a wavelength $\lambda$ of the microwave in the transmission line, $\beta = 2\pi/\lambda$. Using $\alpha\ell\ll 1$ for a high-$Q$ superconducting resonator, Equation (\ref{eq:QWave0Tele}) becomes:
\begin{equation}
    Z_{in}\simeq Z_0 \frac{1-j\alpha\ell\cot{\beta\ell}}{\alpha\ell-j\cot{\beta\ell}}\nonumber
\end{equation}
Near the resonance frequency, $\omega_0$, we can write the frequency $\omega$ as $\omega = \omega_0+\Delta\omega$ so that when $\ell = \lambda/4$ in a quarter-wave resonator, 
\begin{equation}
    \cot{\beta\ell} \simeq -\frac{\pi}{2}\frac{\Delta\omega}{\omega_0}
\end{equation} 
Therefore, \begin{equation}
    Z_{in}\simeq \frac{Z_0}{\alpha\ell+j\pi\Delta\omega/2\omega_0}\label{eq:QWave0}
\end{equation} This equation has the same form as the impedance of a parallel RLC circuit, i.e. $[(1/R_0)+2j\Delta\omega C_0]^{-1}$ with $\omega_0 = 1/\sqrt{L_0C_0}$. By comparing the transmission line resonator with the parallel RLC resonator, we can relate the physical parameters in these two models:
\begin{eqnarray}
R_0 &=& \frac{Z_0}{\alpha\ell}\nonumber\\
C_0 &=& \frac{\pi}{4}\frac{1}{\omega_0Z_0}\end{eqnarray}
The $Q$ factor of the probe superconducting resonator, $Q_0$, is given by Eqn.~\ref{eq:QWave0Q}.

To quantify the kinetic inductance and loss of the sample component using a quarter-wave resonator, we compare Eqn.~\ref{eq:QWave} and \ref{eq:QWave0} to find the frequency shift and change of $Q$ factor by identifying:
\begin{eqnarray}
\alpha\ell &\longrightarrow& \alpha\ell + \frac{r}{Z_0}\\
\frac{\pi}{2}\frac{\Delta\omega}{\omega_0} &\longrightarrow& \frac{\pi}{2}\frac{\Delta\omega}{\omega_0}+\frac{x}{Z_0}\end{eqnarray} where the long arrow indicates a change in physical quantities from those of the pure quarter-wave resonator to those of the quarter-wave resonator connected to ground the plane via the sample component.

\subsubsection{Half-wave resonator}
Similar to the analysis of the quarter-wave resonator, we will first consider the half-wave transmission-line resonator without the sample component. The input impedance is given by Telegrapher equation \cite{Pozar.2005}:
\begin{eqnarray}
Z_{in} &=& Z_0\tanh{(\alpha+j\beta)\ell}\nonumber\\
&\simeq& \frac{Z_0}{\alpha\ell+j\tan{\beta\ell}}\label{eq:HWave0}\end{eqnarray}
Since $\ell = \lambda/2$ in a half-wave resonator, \begin{equation}
    \tan{\beta\ell} \simeq \pi\frac{\Delta\omega}{\omega_0}
\end{equation} Comparing the RLC circuit model, we have:
\begin{eqnarray}
R_0 &=& \frac{Z_0}{\alpha\ell}\nonumber\\
C_0 &=& \frac{\pi}{2}\frac{1}{\omega_0Z_0}\end{eqnarray} The $Q$ factor of the bare probe resonator is given by Eqn.~\ref{eq:Hwave0R0}.

Comparing Eqn.~\ref{eq:HWave} with \ref{eq:HWave0}, 
\begin{eqnarray}
\alpha\ell &\longrightarrow& \alpha\ell + \frac{Z_0}{R}\\
\pi\frac{\Delta\omega}{\omega_0} &\longrightarrow& \pi\frac{\Delta\omega}{\omega_0}-\frac{Z_0}{X}\end{eqnarray} we identify the frequency shift and new $Q$ factor due the sample component.

Table~\ref{tab:RLC} lists the equivalent $R_0$, $L_0$, and $C_0$ in the quarter-wave and half-wave resonators \cite{Pozar.2005}.

\begin{table}[h]
\centering
\begin{tabular}{ c c c c c} 
\hline%\noalign{\hrule height 0.8pt}%\hline
&~~~~~~~~~~~~& Quarter-wave resonator &~~~~~~~~~~~~& Half-wave resonator \\
\hline%\noalign{\hrule height 0.8pt}
$R_0$ & & \parbox{1cm}{\begin{equation}
\frac{Z_0}{\alpha\ell}\nonumber\end{equation}} & & \parbox{1cm}{\begin{equation}
\frac{Z_0}{\alpha\ell}\nonumber\end{equation}} \\
$L_0$ & & \parbox{1cm}{\begin{equation}
\frac{4Z_0}{\pi\omega_0}\nonumber\end{equation}} & & \parbox{1cm}{\begin{equation}
\frac{2Z_0}{\pi\omega_0}\nonumber\end{equation}} \\
$C_0$ & & \parbox{1cm}{\begin{equation}
\frac{\pi}{4\omega_0Z_0}\nonumber\end{equation}} & & \parbox{1cm}{\begin{equation}
\frac{\pi}{2\omega_0Z_0}\nonumber\end{equation}} \\
\hline
\end{tabular}
\caption{\textbf{Formulae of the equivalent $R$, $L$, and $C$ of a transmission-line resonator \cite{Pozar.2005}.}}
\label{tab:RLC}
\end{table}

\subsection{Supplementary information}\label{sec:SI}
Figures of additional information are listed as follow:
\begin{enumerate}
    \item Resonances of two coplanar waveguide resonators which are lithographically identical to quantify fabrication error.
    \item Internal-quality factor $Q_i$ of the resonator with embedded Al sample component from Fig.~\ref{fig:AlfnQ} in the main text measured under various microwave powers output from the microwave source, before any fridge line attenuation.
    \item Internal-quality factor $Q_i$ of a reference resonator versus microwave power from the microwave source at about 20~mK. This resonator is made of Nb without any sample component embedded.
    \item DC electrical transport measurement of the transition temperature of Al, NbSe$_2$, NbN, and Nb used in this experiment.
    \item Schematic of the experimental setup.
    \item Frequency and quality factor temperature dependence data for three separate NbSe$_2$ samples.
    \item Optical image of NbSe$_2$ device including sample, reference, and control resonators.
\end{enumerate}

\begin{figure}[h]
\includegraphics[width=0.9\columnwidth]{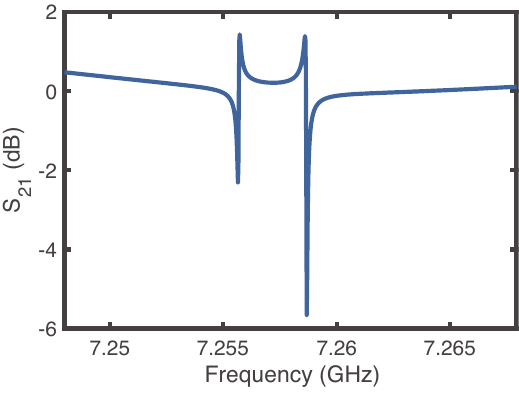}
\caption{Resonances of two coplanar waveguide resonators which are lithographically identical. The difference suggests sub-ppt level uncertainty between the resonance frequency design and actual implementation.}\label{fig:Uncertainty}\end{figure}

\begin{figure}[h]
\includegraphics[width=1\columnwidth]{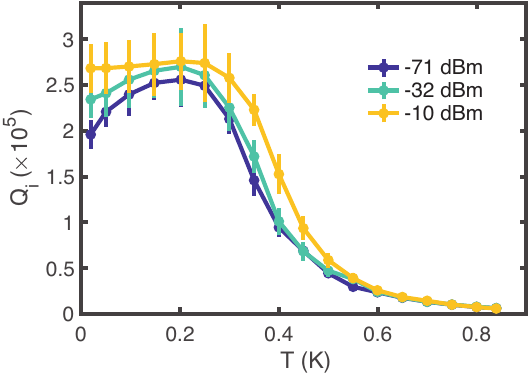}
\caption{Internal-$Q$ factor $Q_i$ of the Al-embedded sample component as a function of temperature at three different powers. The power in the legend denotes the power from the microwave source, before encountering attenuation in the fridge lines. At temperatures well below $\hbar\omega_0/k_B$, $Q_i$ increases with power due to saturation of two-level systems (TLSs) by the measurement power (see Fig.~\ref{fig:TLS}). When the TLSs saturate due to thermal excitations, $Q_i$ has a maximum near $\hbar\omega_0/k_B$, i.e. around 0.2~K for our quarter-wave resonator. $Q_i$ continues to decrease at higher temperatures because of dissipation from the Al sample component due to thermally excited quasiparticles, well described by Mattis-Bardeen theory (see Fig.~\ref{fig:AlfnQ}).}\label{fig:ThreePower}\end{figure}

\begin{figure}[h]
\includegraphics[width=0.9\columnwidth]{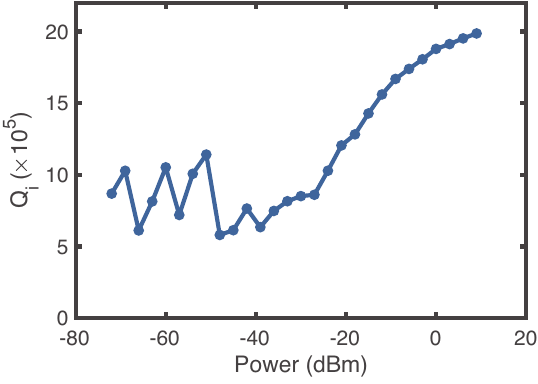}
\caption{Internal-$Q$ factor, $Q_i$, of a reference resonator versus the microwave source power. At high powers, two-level systems (TLSs) are saturated by the power input to measure $S_{21}$, whereas at low powers, TLSs are in their ground state at low temperature, about 20~mK, available to transition to their excited state by absorbing photons from the resonator. Consequently, TLSs provide an additional loss channel through photon dissipation, which degrades $Q_i$. The scattering of $Q_i$ at low power is due to the noise in measuring $S_{21}$ when using a small power. The average number of photons in the resonator is on the order of one in the low-power regime of the plot, evinced by the leveling off of $Q_i$, typical in the single-photon limit.}\label{fig:TLS}\end{figure}

\begin{figure}[h]
\includegraphics[width=0.9\columnwidth]{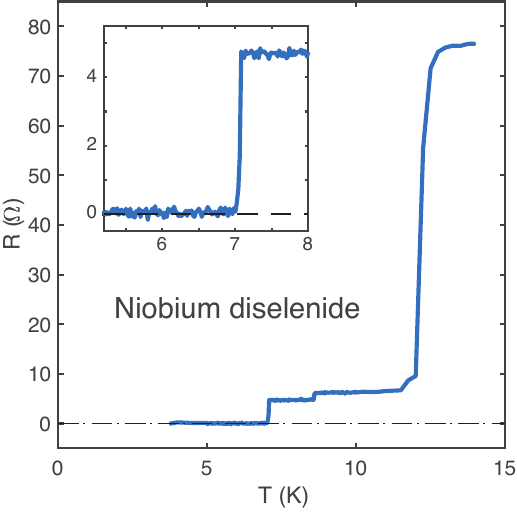}
\caption{DC Resistance of NbSe$_2$, Nb, and NbN versus temperature measured in a NbSe$_2$-embedded resonator (sample component kinetic inductance19) with $T_c$ at 7.06, 8.59, and 12.13~K, respectively. Inset: close-up of the NbSe$_2$ superconducting transition.}\label{fig:TcNbSe2}\end{figure}

\begin{figure}[h]
\includegraphics[width=0.9\columnwidth]{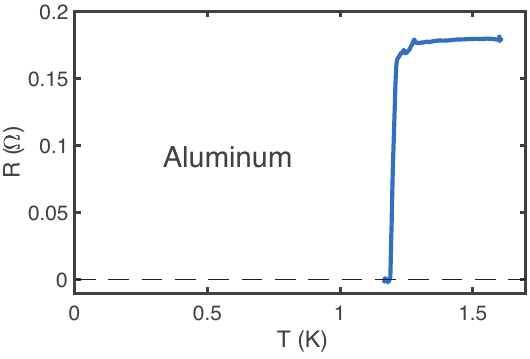}
\caption{Measured $T_c$ of 80~nm-thick Al thin film deposited simultaneously with strip of Al sample component embedded in resonator, from main text.}\label{fig:AlnsTc}\end{figure}

\begin{figure}[h]
\includegraphics[width=0.9\columnwidth]{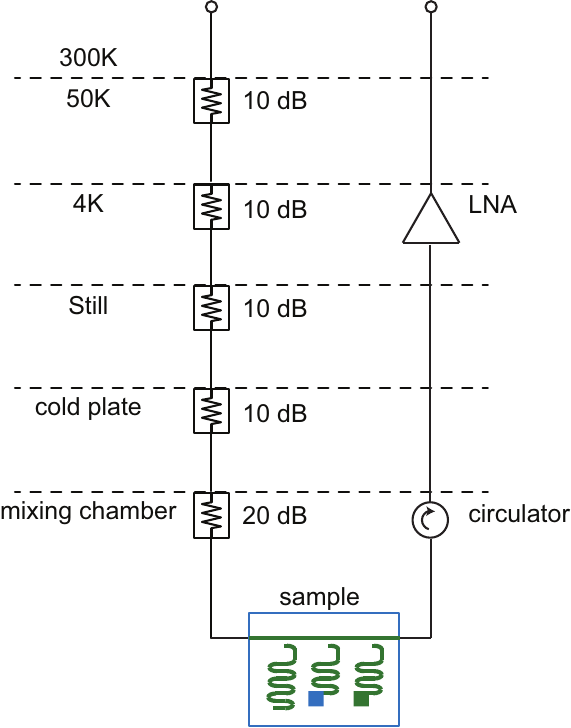}
\caption{Experimental setup of our kinetic inductance measurement. The attenuation of the microwave signal from room temperature to the mixing chamber is typically around 65 dB. A Vector Network Analyzer is used to source and measure.}\label{fig:setup}\end{figure}

\begin{figure*}[t]
\includegraphics[width=1.8\columnwidth]{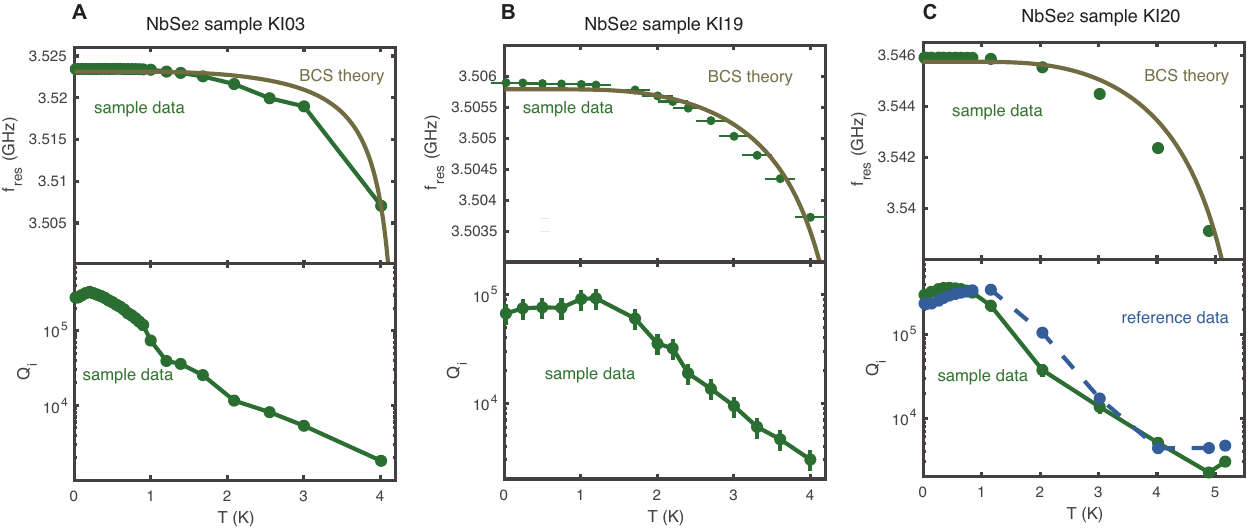}
\caption{Frequency and quality factor data for three different NbSe$_2$ samples.}\label{fig:NbSe2_3samples_fnQ}
\end{figure*}

\begin{figure*}[t]
\includegraphics[width=1.7\columnwidth]{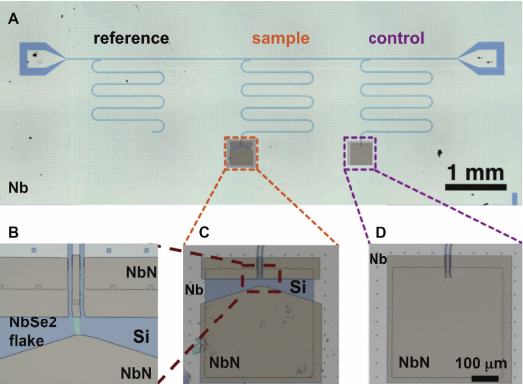}
\caption{Optical image of NbSe$_2$ device made from sputtered niobium and etched into three difference resonators: (A) reference, (B) sample, and (C) control. (A) is composed of niobium and silicon, (B) is a niobium resonator connected to an exfoliated flake of NbSe$_2$ via Ar ion-milling followed by niobium nitride sputtering, and (C) is a niobium resonator with niobium nitride stitching used in (B), with no exfoliated flake.}\label{fig:NbSe2Device}\end{figure*}

%\nocite{*}
\newpage
\bibliographystyle{apsrev4-1}
%\bibliographystyle{naturemag}
%\bibliography{ki,moreRefs}% Produces the bibliography via BibTeX.

\end{document}